\documentstyle[preprint,aps,floats,tighten]{revtex} 

\newcommand{\drawsquare}[2]{\hbox{%
\rule{#2pt}{#1pt}\hskip-#2pt
\rule{#1pt}{#2pt}\hskip-#1pt
\rule[#1pt]{#1pt}{#2pt}}\rule[#1pt]{#2pt}{#2pt}\hskip-#2pt
\rule{#2pt}{#1pt}}
\newcommand{\Yfund}{\raisebox{-.5pt}{\drawsquare{6.5}{0.4}}}
\newcommand{\Yasymm}{\raisebox{-3.5pt}{\drawsquare{6.5}{0.4}}\hskip-6.9pt%
        \raisebox{3pt}{\drawsquare{6.5}{0.4}}}
\newcommand{\Ythreea}{\raisebox{-3.5pt}{\drawsquare{6.5}{0.4}}\hskip-6.9pt%
        \raisebox{3pt}{\drawsquare{6.5}{0.4}}\hskip-6.9pt
        \raisebox{9.5pt}{\drawsquare{6.5}{0.4}}}

\begin{document}


\preprint{\vbox{ \hbox{UCSD/PTH 98-22} \hbox{hep-th/9806079} \hbox{} }}
\title{\bf The Low-Energy Dynamics of  ${\cal N}=1$ SUSY Gauge Theories \\
with Small Matter Content\footnote{Talk presented at the XXXIIIrd
Rencontres de Moriond:  Electroweak Interactions and Unified Theories,
Les Arcs, France, March 14--21 1998.} }

\author{\large Witold Skiba}

\address{\vspace{.2cm} \it Department of Physics,
  University of California San Diego \\ La Jolla, CA 92093--0319, USA}

\maketitle

\begin{abstract}%
We describe the low-energy dynamics of ${\cal N}=1$ supersymmetric gauge
theories with the Dynkin index of matter fields less than or equal to the
Dynkin index of the adjoint plus two. We explain what kinds of
nonperturbative phenomena take place in this class of supersymmetric
gauge theories.

\end{abstract}

\baselineskip=18pt

\section{Introduction}
The knowledge of nonperturbative dynamics of SUSY gauge theories advanced
significantly after the realization of how powerful the restrictions imposed by
holomorphy are~\cite{SW,Seiberg}. It turned out that holomorphy and symmetries
are sufficient to determine the vacuum structure of a large number of
supersymmetric theories. The full Lagrangian of ${\cal N}=1$ theories
contains both holomorphic and non-holomorphic quantities, but only the
holomorphic ones can be determined beyond what is possible to compute
in perturbation theory. 

There are two kinds of holomorphic quantities in ${\cal N}=1$ theories:
the superpotential and the coefficient of the gauge-kinetic term. Symmetry
restrictions and compatibility with various limits are often powerful enough
to determine the exact form on nonperturbatively generated superpotentials.
The gauge-kinetic term is important only when the gauge degrees of freedom
are present in the infrared. When the gauge group is broken or exhibits
confinement, there are no light vector bosons in the spectrum. The field
dependent coefficient of the gauge-kinetic term can be determined  using the
electric-magnetic duality, for which explicit transformations are only
known in the Abelian case. For non-abelian theories we have limited
results about duality. The non-Abelian version of duality relates theories
which flow to the same infrared fixed point and also gives a mapping
between the gauge-invariant operators~\cite{Seiberg2}. So far, the examples of
non-abelian duality are known only in isolated cases, but no general
arguments for finding dual theories have been found. We will not discuss
theories related by such a duality here. On the other hand, ${\cal N}=1$
theories can be studied systematically in a wide range of other
nonperturbative phenomena including confinement, Abelian duality,
quantum deformation of moduli spaces, lifting of the moduli space due
to dynamically-generated superpotentials. We will describe these phenomena
in theories based on simple gauge groups without tree-level superpotentials.

The classical moduli space of supersymmetric gauge theories can be
described by the vacuum expectation values (VEVs) of gauge-invariant
polynomials on the space of chiral superfields. For any given theory, there
is always a minimal set of such polynomials which parameterizes the moduli
space. Some of the operators in the minimal set can be subject to constraints,
which are polynomials in the operators. The choice of the basic set of
gauge invariants is not unique, but the number of such invariants in
the minimal set and the number of constraints among them is unique in a given
theory.

We now describe the restrictions imposed by symmetries on the form of the
dynamically-generated superpotential. The $U(1)$ charges must be chosen so
that the symmetry is free of anomalies. For example, the $R$ charge has to obey
\begin{equation}
  \sum_i (r_i-1) \mu_i + \mu_{adj}=0,
\end{equation}
where $(r_i-1)$ is the R charge of the $i$-th fermion representation,
$\mu_i$ the respective Dynkin index, and $\mu_{adj}$ is the Dynkin index
of the adjoint representation. In our conventions gauginos carry R charge
one. Since we consider theories with no tree-level superpotential, there are
as many non-anomalous global $U(1)$'s as chiral superfields. These symmetries 
restrict the dynamically generated superpotential to be of the form~\cite{ADS}
\begin{equation}
\label{superpot}
  W_{dyn} \propto \left( \frac{\Lambda^{(3 \mu_{adj}-\mu)/2}}{\prod_i
                         \phi_i^{\mu_i}}
                  \right)^{\frac{2}{\mu_{adj}-\mu}},
\end{equation}
where $\mu=\sum_i \mu_i$. Depending on the sign of $\mu-\mu_{adj}$
the characteristic scale of the theory, $\Lambda$, appears in the
numerator or the denominator of the above equation. When $\mu=\mu_{adj}$
a superpotential cannot be generated since all fields carry zero R charge,
while the superpotential must have R charge of two. We will now
discuss nonperturbative phenomena dividing theories according to
the value of $\mu-\mu_{adj}$.

\section{$\mu=\mu_{\lowercase{adj}}+2$}

An example of theory with with $\mu=\mu_{adj}+2$ is supersymmetric $SU(N_C)$
with $N_F=N_C+1$ fields $Q$ in the fundamental representation and $\bar{Q}$
in the antifundamental representation. This theory was found~\cite{Seiberg}
to confine without breaking chiral symmetries. Massless fields of the
confined theory are the mesons $M=Q \bar{Q}$ and the baryons $B=Q^{N_C}$,
$\bar{B}=\bar{Q}^{N_C}$. When $\mu > \mu_{adj}$ the superpotential of
Eq.~(\ref{superpot}) has fields in the numerator and the scale $\Lambda$ in
the denominator. The limit $\Lambda \rightarrow 0$ seems to give a singularity
which is not present in the classical theory. Therefore, most theories with
$\mu > \mu_{adj}$ cannot generate a superpotential. Nevertheless,
supersymmetric QCD with $N_F=N_C+1$ generates a superpotential of the form
\begin{equation}
\label{QCD}
  W=\frac{B M \bar{B} - \det M}{\Lambda^{2 N_C -1}}.
\end{equation}
This superpotential has a consistent classical limit because of constraints
among the mesons and baryons. The equations of motions reproduce precisely
the classical constraints among the $M$'s, $B$'s and $\bar{B}$'s. Thus, the
numerator of the superpotential in Eq.~(\ref{QCD}) vanishes identically
in the classical limit, and the singular behavior is avoided. The moduli
spaces of the classical and quantum theories are identical since the equations
of motion of the quantum theory reproduce the classical constraints.
For example, the point where none of the operators has VEVs belongs
to the quantum moduli space. At this point none of the global symmetries
is broken, and one can verify that the 't~Hooft anomaly matching conditions
are saturated by the mesons and baryons.

It is possible to identify other theories which confine without chiral
symmetry breaking and whose classical and quantum moduli spaces are identical.
Constraints present in the classical theory have to be reproduced
by the equations of motion of the superpotential. All such theories,
termed s-confining, can be found in a systematic manner~\cite{s-conf}.
The confining phase of such theories is continuously connected to
their respective Higgs phase. For VEVs much larger than the characteristic
scale, at generic points on the moduli space of theories with $\mu > \mu_{adj}$
the gauge group is completely broken~\cite{freealg}. Thus the low-energy
degrees of freedom are the fields that are not eaten by the Higgs mechanism.
These are parameterized by the gauge invariants made out of chiral superfields.
Since there is no phase transition between the Higgs and confining phases,
the same fields must comprise the infrared spectrum of the theory near the
origin.

We now summarize the arguments that allow to identify the s-confining
theories~\cite{s-conf}.  First, the confining and Higgs phases are
indistinguishable whenever sources in all possible representations can be
screened by massless dynamical fields. This requires that an s-confining
theory has at least one field in a faithful representation. Second, the
superpotential must be smooth at the origin. Comparison with
Eq.~(\ref{superpot}) implies that $\mu=\mu_{adj}+2$.
A singularity of the superpotential would indicate that additional
massless states are present. These two conditions limit the number
of candidate theories to a relatively short list. The last criterion
also comes form the fact that the same massless fields are present on
the entire moduli space. Giving VEVs to some of the fields, such that
the gauge group is only partially broken, leads to new effective theories.
But these new theories should also be described
in terms of gauge invariant operators alone. In other words, an s-confining
theory has to flow to an s-confining theory. Let us see how this works in
practice. For example an $SU(4)$ theory with 5 $\Yasymm$, obeys the index
restriction $\mu=\mu_{adj}+2$. However the two-index antisymmetric
representation is not faithful, therefore this theory cannot be s-confining.
As another example consider $SU(4)$ with $\Yasymm + 2 \, (\Yfund +
\overline{\Yfund})$. Along a certain flat direction this theory flows to
$SU(2)$ with $8 \, \Yfund$. An $SU(2)$ theory with $8 \, \Yfund$
is not s-confining as it can be described in terms of a dual gauge
theory~\cite{Seiberg2}. Therefore $SU(4)$ with $\Yasymm
+ 2 \, (\Yfund + \overline{\Yfund})$ is not s-confining.

Using the criteria we have just described it is possible to analyze all
theories with $\mu=\mu_{adj}+2$ and eliminate the ones which are not
s-confining. For the remaining candidate theories the low energy degrees of
freedom were found, and these fields indeed satisfy 't~Hooft anomaly matching
conditions at the origin of the moduli space~\cite{s-conf}.

\section{$\mu=\mu_{\lowercase{adj}}$}

In the previous section we have shown how to identify s-confining theories.
Adding a mass term  for a flavor (or any other real representation with
$\mu_R=2$) yields theories with $\mu=\mu_{adj}$. Decoupling the massive fields
gives theories with a smaller field content. For example, integrating out
a flavor from s-confining supersymmetric QCD, one obtains a theory with
$N_F=N_C$.  As we already mentioned when $\mu=\mu_{adj}$ a dynamical
superpotential can not be generated. The classical moduli space of
supersymmetric QCD with $N_F=N_C$ has one constraint among the gauge
invariants, which is $(\det M - B \bar{B})=0$. The quantum moduli space turns
out to be different from the classical one, and it is described by the
constraint $(\det M - B \bar{B})=\Lambda^{2 N_C}$.
The modified constraint forces at least one combination of fields to have
non-zero VEVs. If $\langle \det M \rangle \neq 0$, then non-abelian flavor 
symmetries are broken; if $\langle B \bar{B} \rangle \neq 0$ then the baryon
number is broken. For small VEVs SUSY QCD with $N_F=N_C$ is confining, while
for large VEVs it is in the Higgs phase. Because the fundamental representation
is faithful there is no phase transition between the two phases. One can check
that the same happens when integrating out a flavor from any other s-confining
theory.

Not all $\mu=\mu_{adj}$ theories can be obtain by integrating out fields from
s-confining theories. Nevertheless, the infrared dynamics of all
other theories with  $\mu=\mu_{adj}$ has been determined~\cite{qdf,coulomb}.
$\mu=\mu_{adj}$ theories which have constraints among the basic invariants
that parameterize the classical moduli space have low-energy dynamics similar
to that of SUSY QCD with $N_F=N_C$. The set of
operators and appropriate quantum modification of the moduli space space have
been found in Ref.~\cite{qdf}. In some cases the modification is field
dependent, that is, a classical constraint picks up a modification proportional
to the scale of the theory times a product of fields. The low-energy dynamics
is the same irrespective of the form of the quantum constraint. Near the
origin of the moduli space the theory is confining, far away it is Higgsed,
and there is no phase transition between the phases.

Indeed, all theories with $\mu=\mu_{adj}$ and constraints among the basic
gauge invariants have fields in a faithful representation.
Such theories have completely broken
gauge groups at large generic VEVs. $\mu=\mu_{adj}$ theories which do not
have constraints are all theories with a single adjoint superfield and three
other examples~\cite{coulomb}. Theories with an adjoint superfield
automatically have ${\cal N}=2$ supersymmetry; we will not discuss them here.
The remaining three ${\cal N}=1$ examples are $SO(N)$ with $(N-2)\, \Yfund$,
$SU(6)$ with $2 \, \Ythreea$,
and $Sp(6)$ with $2\, \Yasymm$. None of these theories contains fields
in a faithful representation. Generic VEVs of these theories break the gauge
group to a single $U(1)$ or a product of $U(1)$'s. It is therefore possible
to determine the coefficient of the gauge-kinetic term using
electric-magnetic duality~\cite{SW,phases}.
The auxiliary Seiberg-Witten curves which encode information about the
holomorphic coefficient of the gauge-kinetic term have been found for these
theories~\cite{SO,coulomb}. It turns out that the $U(1)$ gauge bosons are
present at every point of the moduli spaces.

It is interesting that these three examples exhaust the list of theories
in the Abelian Coulomb phase everywhere on the moduli space~\cite{coulomb}.
Since the Seiberg-Witten curve is written in terms of chiral gauge-invariant
operators and the scale $\Lambda$, one can argue based on R symmetry
considerations that $\mu$ must equal $\mu_{adj}$. It is easy to understand
that there cannot be $U(1)$ symmetries
for generic VEVs when $\mu > \mu_{adj}$ since the gauge is broken completely
in such cases. For $\mu < \mu_{adj}$, as we will explain in the next section,
all theories have dynamically-generated superpotentials which lift the moduli
space. In a few cases theories have multiple branches: a branch with 
a superpotentials and a branch with zero superpotential and confinement near
the origin. However, none of the theories with $\mu < \mu_{adj}$ has Abelian
factors in the low-energy spectrum.

\section{$\mu<\mu_{\lowercase{adj}}$}

Since the description of all theories with $\mu=\mu_{adj}$ is known, one could
in principle integrate out fields in the real representations and obtain the
description of theories with $\mu<\mu_{adj}$. Integrating out can sometimes
be very difficult due to cumbersome algebra. There are, however, general
arguments which allow one to classify all $\mu<\mu_{adj}$ theories and
explain their dynamics~\cite{freealg}. First of all, each theory with
$\mu<\mu_{adj}$ has an unconstrained set of basic gauge invariants. At the
quantum level, such theories turn out to always have a branch with a dynamical
superpotential described in Eq.~(\ref{superpot}). Such a superpotential forces
fields to infinite VEVs and lifts the moduli space. Some theories have an
additional branch on which there is no
superpotential~\cite{s-conf,SO,Sp,GA-PRL}.
One can understand the multiple branch structure and why all theories have
branches with non-zero superpotentials by examining the smallest unbroken
subgroup on the moduli space~\cite{freealg}.

There are two known mechanisms for generating dynamical superpotentials:
instantons and gaugino condensation. When the gauge group is completely broken,
one-instanton contributions generate a superpotential. It is possible to
connect all such theories to an $SU(2)$ theory with one flavor,
in which an explicit
instanton computation was carried out~\cite{ADS}. All theories with a
completely broken gauge group flow along a flat direction to an $SU(2)$.
Instantons in that subgroup can be related to a nonvanishing
superpotential in the original theory.

Gaugino condensation takes place in SUSY Yang-Mills theories with arbitrary
gauge groups. Whenever
an unbroken subgroup remains at generic values of the moduli fields, the
gaugino condensate contributes to the superpotential. If the unbroken subgroup
is simple, then gaugino condensation can be related by scale matching to a
nonvanishing superpotential of the original theory. When the unbroken subgroup
is semi-simple, gaugino condensations from each factor contribute to the
superpotential. Depending on the relationship between the scales of
different factors, these contributions do or do not cancel. If the
contributions do not cancel, the original theory has only branches with
non-zero superpotential. Otherwise one obtains theories with both zero
and non-zero branches~\cite{freealg}.

\section{Conclusions}
The vacuum structure of s-confining theories, which are a large subset of
the $\mu=\mu_{adj}+2$ list, and of all theories with $\mu \leq \mu_{adj}$
has been determined. The nonperturbative phenomena exhibited by those
theories are confinement with or without chiral symmetry breaking,
Abelian Coulomb phase, and dynamical generation of superpotentials. 
Such theories with a small matter content provide the basic blocks for
supersymmetric model building. The majority of model building efforts so far,
including compositeness and dynamical supersymmetry breaking, use theories
of this kind. It would be very interesting to find out about the dynamics
of the remaining asymptotically-free supersymmetric theories with
$\mu > \mu_{adj}$.

\acknowledgements

I am grateful to Csaba Cs\'aki, Gustavo Dotti, Aneesh Manohar, and Martin
Schmaltz for fruitful collaborations on the topics discussed in this talk.
I thank  Csaba Cs\'aki and Aneesh Manohar for comments on the manuscript.
This work was supported in part by the U.S.\ Department of Energy under grant
no.\ DOE-FG03-97ER40506.


\begin{references}

\bibitem{SW} N. Seiberg and E. Witten, Nucl.\ Phys.\  {\bf B426} (1994) 19;
Nucl.\ Phys.\  {\bf B431} (1994) 484.

\bibitem{Seiberg} N. Seiberg, Phys.\ Rev.\ {\bf D49} (1994) 6857.

\bibitem{Seiberg2} N. Seiberg, Nucl.\ Phys.\  {\bf B435} (1995) 129.

\bibitem{ADS} I. Affleck, M. Dine, and N. Seiberg,
Phys.\ Rev.\ Lett.\ {\bf 51} (1983) 1026; \\
Nucl.\ Phys.\ {\bf B241} (1984) 493.

\bibitem{s-conf} C. Cs\'aki, M. Schmaltz, and W. Skiba,
Phys.\ Rev.\ Lett.\ {\bf 78} (1997) 799; \\
Phys.\ Rev.\  {\bf D55} (1997) 7840.

\bibitem{freealg} G. Dotti, A. Manohar, and W. Skiba, hep-th/9803087.

\bibitem{qdf} B. Grinstein and D. Nolte, Phys.\ Rev.\ {\bf D57} (1998) 6471;
hep-th/9803139; \\ P. Cho, Phys.\ Rev.\ {\bf D57} (1998) 5214.

\bibitem{coulomb} C. Cs\'aki and W. Skiba, hep-th/9801173.

\bibitem{phases} K. Intriligator and N. Seiberg, 
Nucl.\ Phys.\  {\bf B431} (1994) 551.

\bibitem{SO}  K. Intriligator and N. Seiberg, 
Nucl.\ Phys.\  {\bf B444} (1995) 125.

\bibitem{Sp} P. Cho and P. Kraus, Phys.\ Rev.\ {\bf D54} (1996) 7640; \\
C. Cs\'aki, W. Skiba, and M. Schmaltz, Nucl.\ Phys.\ {\bf B487} (1997) 128.


\bibitem{GA-PRL} G. Dotti and A. Manohar,
Phys.\ Rev.\ Lett.\ {\bf 80} (1998) 2758.




\end{references}
\end{document}